# CIRPASS: description, performance and astronomical results

Ian Parry[a*], Andrew Bunker[b], Andrew Dean[a], Michelle Doherty[a], Anthony Horton[a], David King[a], Marie Lemoine-Busserole[a], Craig Mackay[a], Richard McMahon[a], Steve Medlen[a], Rob Sharp[c], Joanna Smith[a].

[a]Cambridge University, Institute of Astronomy, Madingley Rd, Cambridge, CB3 0HA, UK;
[b]University of Exeter, School of Physics, Stocker Road, Exeter, EX4 4QL, UK;
[c]Anglo-Australian Observatory, PO Box 296, Epping, NSW 1710, Australia.

## ABSTRACT

The Cambridge Infra-red Panoramic Survey Spectrograph (CIRPASS) is described. This near-infrared (NIR) spectrograph has been used on the 8m Gemini-South Telescope, the 3.9m Anglo-Australian Telescope (AAT) and the 4.2m William Herschel Telescope (WHT). Its performance in both integral field mode and multi-object mode is discussed and some scientific highlights are presented. A multi-IFU system, which is currently under construction, is also described.

## 1. INTRODUCTION

CIRPASS is a near-infrared, fibre-fed spectrograph. It is a private instrument built by the Institute of Astronomy in Cambridge UK for use on 4-10m class telescopes. To achieve versatility, both in terms of telescope choice and mode of operation, it is fibre-fed. The operating wavelength range is 0.9-1.8 microns (the use of silica fibres longer than 1 metre rules out operation in the K-band). This is extremely well suited to our main scientific goal, which is the study of high redshift galaxies by observing emission lines from star-forming regions in the very dark gaps between the telluric OH lines. CIRPASS is funded via PPARC and a donation from the Raymond and Beverly Sackler Foundation.

CIRPASS offers ultra-sensitive, multi-object spectroscopy (MOS), multi-IFU spectroscopy, standard integral field spectroscopy (IFS) and low to medium-resolution spectro-photometry. Its powerful capabilities are not generally available via the current and planned facility instruments on 8-10m telescopes. CIRPASS was the first NIR IFU to be used on an 8-m class telescope and we are the first group to publish results obtained with a fibre-fed NIR multi-object spectrometer. CIRPASS was the second instrument to use a Hawaii-II detector on a telescope. A very detailed Ph.D. thesis[1] gives a full technical description of the CIRPASS spectrograph and its IFU mode.

## 2. DESCRIPTION AND PERFORMANCE

### 2.1. Spectrograph

The whole spectrograph system is operated inside a heavily insulated chamber, which is cooled internally to a temperature selected in the range –44 < T < –40 deg Celsius. A single stage refrigeration system is used. The internal temperature is controlled to 0.1 deg making the spectrograph extremely stable. It takes approximately 24 hours to completely cool the internal structures to a stable temperature. Water vapour inside the chamber condenses on the evaporator coils and not on the optical surfaces, which are warmer. Using active heating, the inside can be warmed up to ambient temperature in approximately 4 hours. Figure 1 shows CIRPASS on Gemini-South.

The optical layout of the CIRPASS spectrograph is shown in Figure 2. Being fibre fed, it is floor mounted and is very stable because it experiences a constant gravity vector. The design has an intermediate spectral focal plane (at the mask mirror) for two reasons: firstly it allows the option of suppressing the telluric OH lines in hardware, secondly it allows a near-Littrow grating configuration so that high spectral resolutions can be obtained via steep grating angles. In practice only the second of these two reasons is used – the OH emission is simply avoided in the data reduction process. The final part of the system is a cryogenic camera, which images the light on to a Rockwell Hawaii-II array. The detector in this camera is cooled using solid nitrogen (we pump on the LN2 can). The detector temperature is kept stable by actively heating it. The operating temperature can be set anywhere in the range 68K-90K. Typically we operate it at 70K.

---

[*] irp@ast.cam.ac.uk, phone +44 1223 337092, fax +44 1223 337523

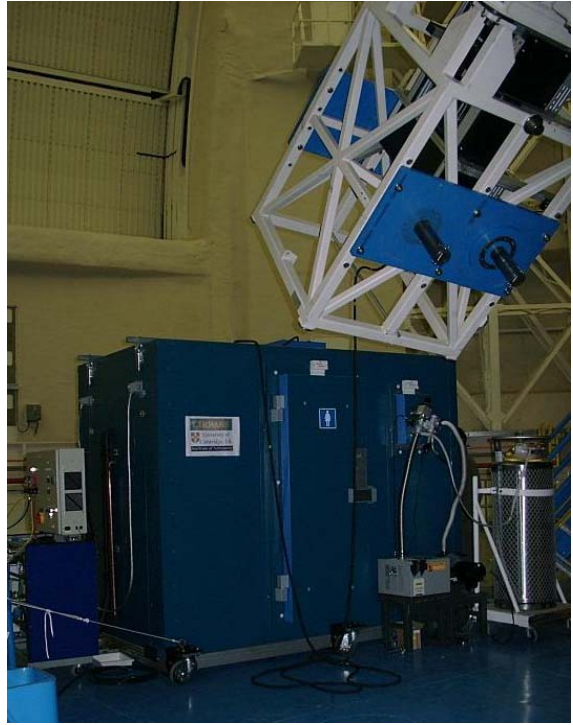

Fig 1. CIRPASS on Gemini-S in Aug 2002. The large blue box on the floor is the CIRPASS cold room. The structure to the top right is a ballast weight assembly containing the CIRPASS IFU. The fibre feed can be seen connecting the two.

The first part of the system is an off-axis Schmidt in double-pass, the first pass acting as the collimator and the second acting as a camera. The light from the fibres is collected by the f/5.5 collimator. The fibre slit is designed so that each fibre aims its output beam directly at the spectrograph entrance pupil. For observations at the faintest limits, a reflection grating working in first order provides spectra on the mask at a spectral resolution which is high enough (R>3100 at 1.3µm) to work between the OH lines. The beam size at the grating is 150mm in diameter allowing standard, large Richardson Labs gratings to be used. The f/5.5 spectrum at the mask mirror is physically reduced to the size of the HAWAII-II array by a third pass through the Schmidt (which collimates the beam) and an f/1.35 camera. To select the central wavelength we tilt the grating. We have gratings with 400, 600, 830 and 1200 l/mm (the 600 and 1200 gratings are kindly on loan from the AAO). Combined with the IFU feed and the MOS feed (which have different effective slit widths) these give a variety of spectral resolutions in the range 2,500<R<12,000. We plan to get an echelle grating and a low resolution grating to extend the resolution range to 1,000<R<26,000. For MOS work we have used the 830 l/mm grating and for IFU work we have used the 400 l/mm grating.

The camera delivers light to the Hawaii-II detector at f/1.35. The large beam-size, the fast f-ratio and the large format detector made the design of this camera quite a challenge. To avoid the use of aspheric surfaces and expensive materials such as calcium fluoride and barium fluoride a design with no correction for chromatic aberration was adopted. The ensuing variation of focal length with wavelength is dealt with by tilting the detector. The detector has a 3-point kinematic mount, which provides full tip, tilt and piston motions. The camera has 5 lenses made of standard Schott glasses (CIRPASS does not operate in the K-band) the largest of which is 224mm in diameter.

A filter which blocks light longward of 1.85 microns is permanently in the beam between the last camera lens and the detector and is cooled to the same temperature as the detector. Blocking filters which block longward of 1.4 and 1.67 microns are also available in the camera filter wheels. To keep the thermal background from the instrument (mainly from the collimator part which operates at –42 deg Celsius) below that of the detector dark current and the inter-line sky one of these filters is used in conjunction with the 1.85 micron blocking filter.

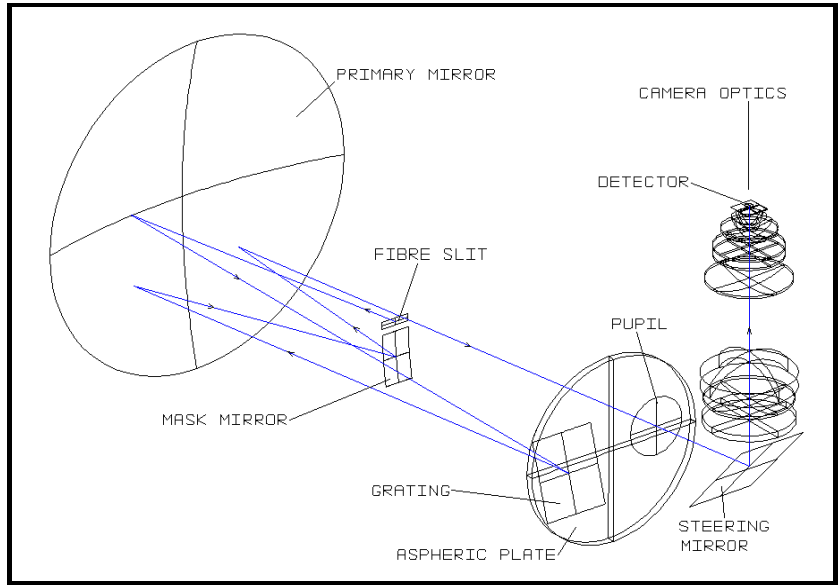

Figure 2: the layout of the CIRPASS spectrograph optics

## 2.2. The Integral field Unit

We use a lens array in the image plane and each lens in the array forms a micro-pupil image. A single fibre behind each lens is used to collect the light. To get a lens array of the highest possible quality we made our own by tiling glass hexagonal lenses on a single glass substrate. The hexagonal lenses were ground and polished individually using conventional methods. Each lens in the array is 3mm across the diagonals for ease of manufacture. The advantage of this approach is that the fill factor is excellent (>95%), the light entering the fibre is telecentric, there is minimal cross-talk between lenses and the whole array was easily AR coated. Figure 3 shows the completed IFU and its geometry on the sky.

The lens array is far too big to match the telescope focal plane image scale directly even on the largest telescopes, so some fore-optics are required to produce a suitably magnified focal plane. This means we can use our IFU on many different telescopes and with many image scale options simply by changing the small magnifying lens at the front of the system. Figure 4 shows how the fore-optics are arranged.

To ensure that the 104 micron diameter fibres on the back of the array are aligned to within 2 microns of the pupil image each fibre was individually positioned and epoxied while steering a back-illuminated fibre image on to the common pupil image formed by the field lens. There are 490 fibres behind the lens array.

## 2.3. Multi-object fibre feed

This is a plug-plate system with 150 individual fibres of 250 micron core diameter. When properly matched to the telescope f-ratio they give 2 arcsec apertures on a 4m telescope and 1 arcsec apertures on an 8m telescope. The full FOV of the telescope can be used. The first three runs (two on the AAT and one on the WHT) with this feed used the AAO's FOCAP system, which had been decommissioned and stored at the AAO for about 10 years. For future runs we will use a plug-plate system of our own construction. Fibre bundles, which plug into the same plate and operate at optical wavelengths feed a fast readout cooled CCD for acquisition and guiding. Because the fibres have a larger diameter than the IFU ones a higher dispersion grating (830 l/mm) is used to work in clean regions between the OH lines.

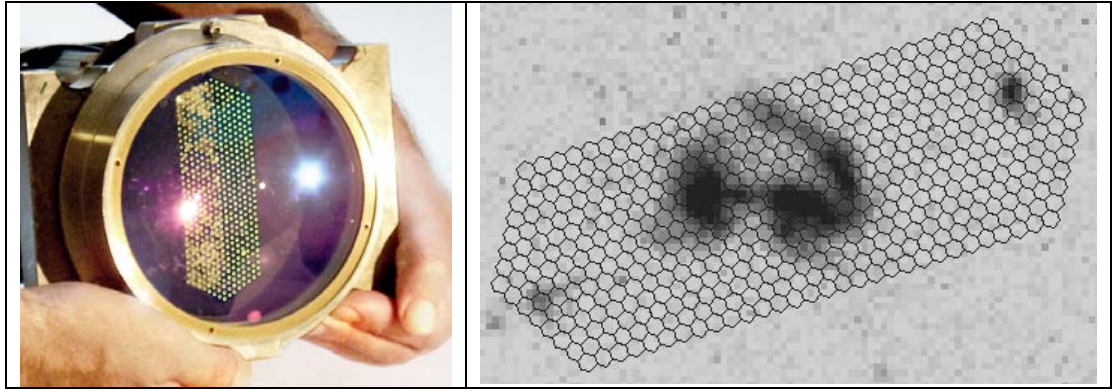

Fig 3. Left: the CIRPASS IFU with its field lens. An image of the bright light illuminating the IFU can be seen reflected in each lens of the array. The fill factor of the lens array is >95%. Right: the geometry of the CIRPASS IFU. In this example the lenses are 0.25 arcsec across and the FOV is 10×4 arcsec. Scales from 0.7 to 0.05 arcsec per lens (with corresponding changes to the FOV) are possible on 4-8m telescopes by changing the choice of fore-optics lens.

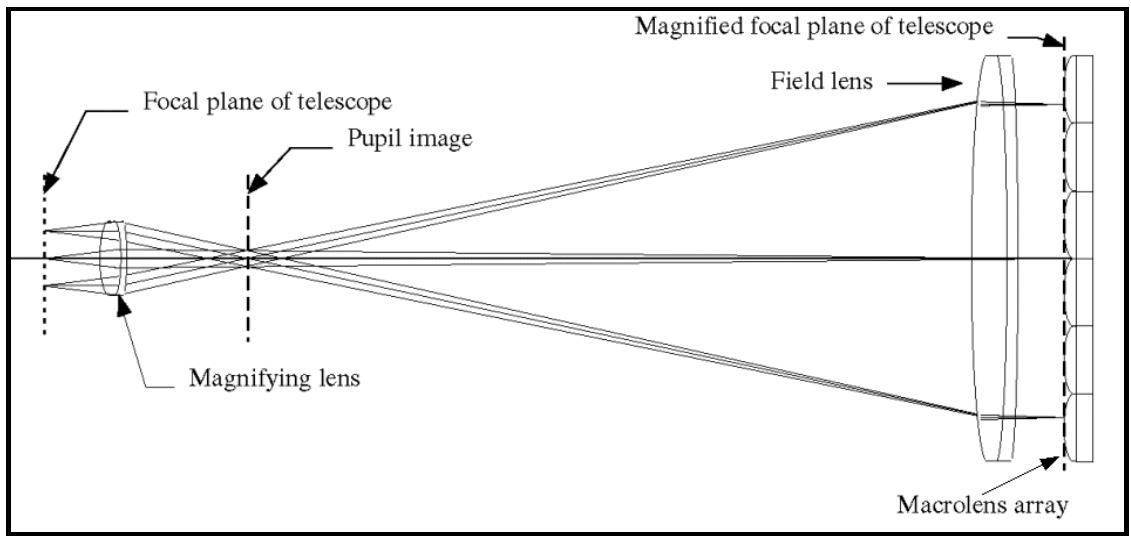

Fig 4. The CIRPASS IFU fore-optics. Choosing the appropriate magnifying lens sets the image scale on the lens array. There is an optical fibre (not shown) at the centre of the focal plane of each lens in the array. The pupil image formed by the magnifying lens is imaged onto the face of each fibre.

**2.4. Performance**

The detector dark current (measured using a metal plate in the cryogenic filter wheel) is 0.05 e/sec/pixel. When the metal plate is removed and the detector views the interior of the cold room through the blocking filters, the total measured background (which includes the detector dark current) is 0.1 e/sec/pixel. The readnoise is 20 e rms per single read. The effective readnoise for a single exposure however is 10 e rms because we use multiple non-destructive reads.

The total measured throughput (photons available in a full 8m aperture above the Earth's atmosphere compared to photons recorded in data file) in the IFU mode on Gemini South was 5%.

On Gemini with the IFU we measured line fluxes of $1.5\times10^{-17}$ erg s$^{-1}$ cm$^{-2}$ in 4 hours (10 sigma). On the WHT and the AAT we could detect emission lines as faint as $5\times10^{-17}$ erg s$^{-1}$ cm$^{-2}$ in 3 hours (5 sigma). On an 8m telescope the MOS mode should be able to detect lines as faint as $6\times10^{-18}$ erg s$^{-1}$ cm$^{-2}$ in 10 hours (5 sigma).

## 3. OBSERVING PROGRAMS

At the time of writing CIRPASS has been used in IFU mode on Gemini South and in MOS mode at the AAT and the WHT. Table 1 gives details of the 6 runs. The first run on Gemini was to demonstrate the power of IFUs via a range of observing programs. Following this successful run the Gemini community was invited to apply for time on Gemini South with CIRPASS. A total of 222 hours was awarded for queue-scheduled CIRPASS observations in two separate runs. The MOS runs at the AAT and the WHT have been successful but have been badly compromised by the weather, especially the second AAT run which was almost totally wiped out. The detector was changed from a Hawaii-I to a Hawaii-II in Sept 2003, just before the start of the second AAT run. Results from these runs are given in the following sections. A further 4 night MOS run on the WHT is scheduled for Oct 2004.

CIRPASS is a large instrument. The spectrograph weighs approximately 1.3 tonnes and when it is moved from one observatory to another it travels as 14 cases of various sizes forming a total shipment of around 3 tonnes. Most of the work to prepare CIRPASS for an observing run (and un-installing it at the end) is done by Cambridge staff. Although it is big, its size does not particularly add to the amount of support effort required from the local observatory staff. Working on very faint sources in between the OH lines means we have very long exposure times. There has therefore never been a need to have the CIRPASS control software fully integrated with the TCS. At Gemini, our software communicated with the TCS to get header information because the data was destined for the Gemini archives. There was no communication between the CIRPASS system and the TCS at the WHT or the AAT. This all worked very well and avoided unnecessary work by observatory staff.

| Telescope | date | mode | Number nights Allocated/clear | Science program |
|---|---|---|---|---|
| Gemini South | Aug 2002 | IFU | 10/8 | Demo science program (PIs, Bunker, Davies and Gilmore) |
| AAT | Oct 2002 | MOS | 8/7 | z~1 galaxies |
| Gemini South | Mar 2003 | IFU | 16/14 | Queue (14 programs) |
| Gemini South | July 2003 | IFU | 6/2 | Queue (continued) |
| AAT | Oct 2003 | MOS | 8/1 | z~1 galaxies, BDs in sigma Orionis |
| WHT | Dec2003/Jan 2004 | MOS | 12/6 | z~1 galaxies, trapezium cluster BDs |
| WHT (awarded) | Oct 2004 | MOS | 4/? | z~1 galaxies |

Table 1: CIRPASS observing runs

# 4. INTEGRAL FIELD SPECTROSCOPY

## 4.1. CFRS 22.1313

This galaxy with a known redshift of 0.82 was observed as part of the demo science program. We were able to measure the star-formation rate and the internal dynamics from H-alpha emission. We find[2] a velocity separation of 220 ± 10 km/s for the two H-alpha emission regions (see figure 5). The lower limit on the rotational velocity is consistent with no evolution of the rest B-band Tully Fisher relation and with disc brightening of no more than ~1 mag at z=0.8. Integral field spectroscopy overcomes the inherent uncertainties of long slit spectroscopy – missing light on the slit jaws and possible misalignment of the slit may lead to an erroneous estimate of the rotation velocity.

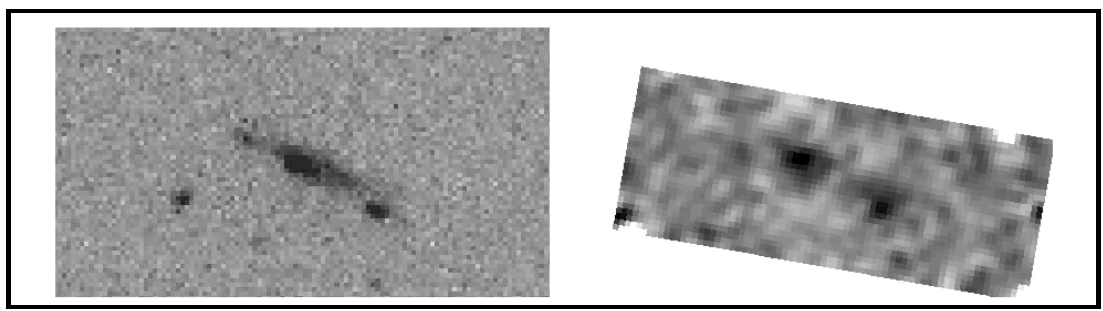

Fig 5. Left: a 6 × 11 arcsec part of the HST F814W image of CFRS 22.1313 which corresponds to the B-band rest frame image at z=0.8. The two prominent HII regions are clearly visible against the extended highly inclined galactic disc. Right: these regions are seen in our CIRPASS spectra. We show a 2D slice from our 3D data cube corresponding to the H-alpha emission. The spatial scale and orientation are the same as in the HST image

## 4.2. The Einstein cross Q2237+0305

This z=1.69 QSO appears as a quadruple image due to a foreground lensing galaxy at redshift z=0.03. Because the broad lines in the QSO spectrum come from the small central region of the QSO and the narrow lines come from the much larger surrounding region it is possible to put constraints on the mass substructure in the fore-ground lensing galaxy by comparing the flux ratios of the broad and narrow lines. This is because the observed ratios relate directly to the areas (unresolved) of the broad and narrow line regions as produced in all four images by the lensing galaxy. It is crucial that all the flux in the narrow and broad lines is collected simultaneously for all four images. A slit spectrograph could not make this observation. Furthermore, as the QSO is at high redshift the emission lines have to be observed in the NIR. This was therefore an ideal project for CIRPASS. The lines chosen were the broad H-beta line and the adjacent narrow [OIII] 5007 line (see figure 6). Our observations[3] provided evidence for substantial mass substructure in the lensing galaxy on scales of ~$10^7$ solar-masses – even more than predicted by the CDM models that gave rise to the so-called "missing dwarf problem" for our own galaxy.

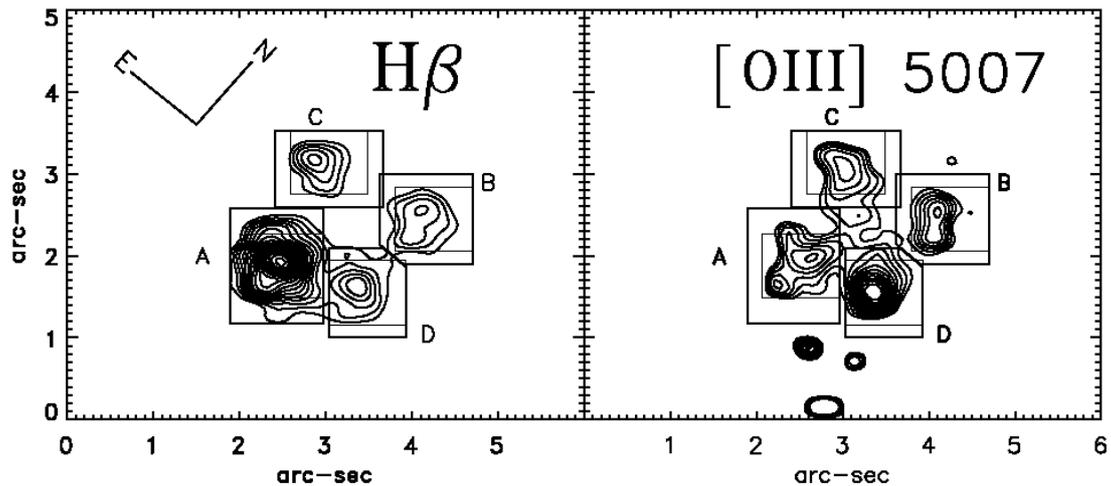

Fig 6. Maps of the H-beta and [OIII] emission of Q2237+0305as obtained with CIRPASS. Note how different they look showing that the amplifications of the broad and narrow line regions are quite different.

### 4.3. The central regions of M83

A NIR IFU is a powerful tool for penetrating the dust seen at optical wavelengths in the centres of nearby galaxies. We observed the central regions of the nearby starburst galaxy M83 in both the J and H bands as part of the queue-scheduled program on Gemini-South. The data obtained[4] contains a wealth of information on super-novae activity, star-formation rates and internal dynamics. What we were particularly interested in doing was to produce an age-map of the recent star formation by comparing the strengths of the CO absorption in the giants with the Pa-beta emission from the HII regions. Figure 7 shows the age map we produced showing evidence of an age-gradient across the field.

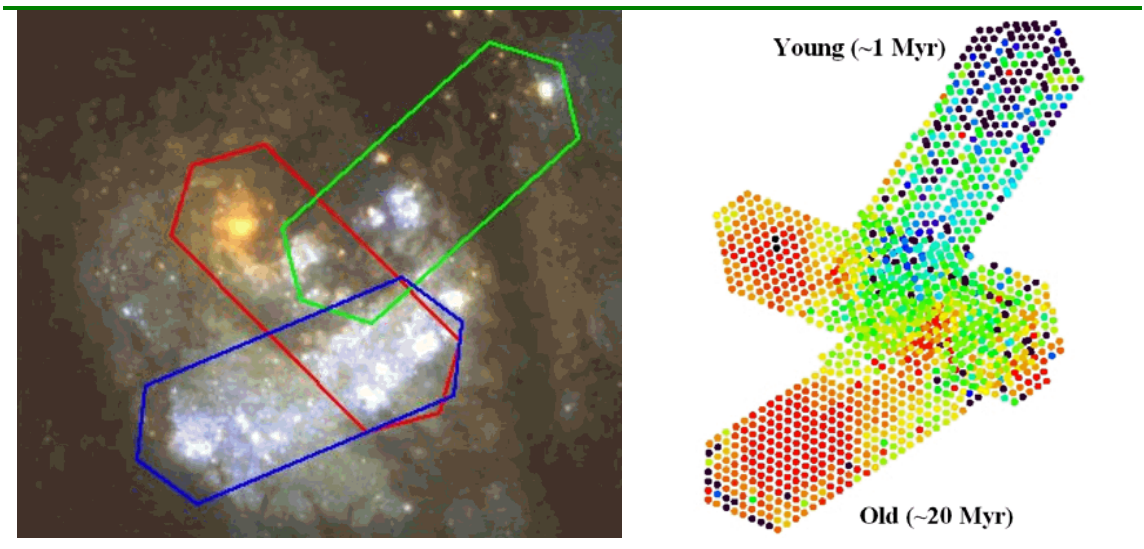

Fig 7. Left: an image of M83 (from Harris et al, 2001, AJ 122 3046) showing how the CIRPASS IFU was deployed on Gemini-S via three pointings. Right: the age map derived by taking the ratio of the Pa-beta emission line strength to the CO (1.623 micron) absorption line strength. A clear age gradient is seen with the youngest stars at the top right associated with the dark dust lane in the optical image.

# 5. MULTI-OBJECT SPECTROSCOPY

We have used our MOS fibre feed for 2 programs: an investigation of the low mass brown dwarfs in young nearby clusters and a spectroscopic survey to accurately measure the star-formation rate at z~1. Here we report very briefly on the second of these two. Both have so far been hampered by poor weather. They are also quite a challenge on 4-m class telescopes.

In our first run at the AAT our galaxy samples were mostly selected from photometric redshifts. We got several detections of H-alpha emission in the Chandra Deep Field South (CDFS) in relatively short exposures (see figure 8). As for the IFU mode, we also use a beam-switching technique for the MOS mode. 75 galaxies are observed at a time. Two fibres are assigned to each galaxy – one on the object and the other on sky. The telescope is nodded between the two positions so that the galaxies are either going down the A-position fibres or the B-position fibres. The 2D data frames can be subtracted from each other directly before spectral extraction. Because the sky spectrum that is subtracted off has been observed with exactly the same instrument set up as the galaxy, the sky subtraction has no systematic errors (systematic errors associated with subtracting sky measured with different fibres have led many astronomers to wrongly claim that good sky subtraction is impossible with fibres).

For the second MOS run at the AAT we had a much better selection technique – known redshifts from optical spectroscopy were used to select galaxies where the H-alpha line would fall in between the OH lines. Unfortunately, this run was hit by very bad weather. Our third run at the WHT had reasonable weather and we were able to measure H-alpha fluxes[5] in 7 galaxies in the Hubble Deep Field North (HDFN) in an exposure when we had good conditions (see figure 9). We find that the star formation rate is ~3 times higher than that measured using the UV continuum at z~1. By coadding all our spectra after shifting them to the rest frame we can get the spectrum of an "average" galaxy at z~1 and hence get a very good estimate of the global SFR. This shows the real power of multi-object spectroscopy which effectively makes our 4-m telescope as powerful as a 30-m telescope studying a single z~1 galaxy!

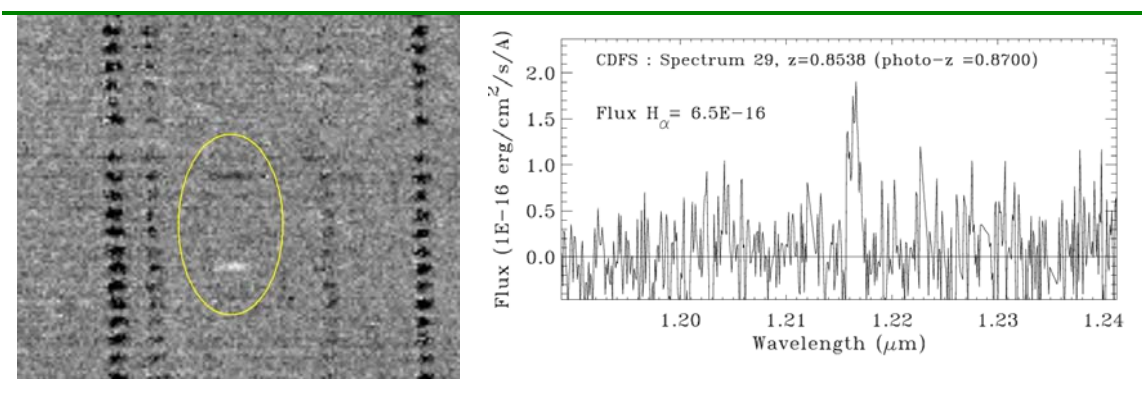

Fig 8. Detections of H-alpha emission, in a one hour CIRPASS exposure, for a galaxy in the Chandra Deep Field South from the first AAT MOS run. The left-hand panel shows the beam-switched spectra on the detector (the wavelength direction is left-to-right). The H-alpha line appears as a positive and negative pair because the galaxy was switched between two fibres. This technique gives excellent continuum sky-subtraction (the OH residuals are because the OH intensity is extremely variable). The right-hand panel shows the fluxed spectrum.

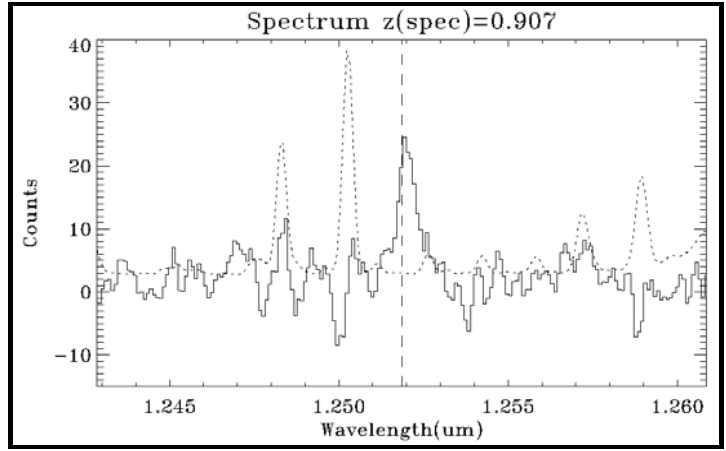

Fig 9. An example of a spectrum for one of the detections of a z~1 galaxy in the Hubble Deep Field North from our WHT CIRPASS MOS observing run in Jan 2004. A sky spectrum is overlaid as a dotted line. The expected position for H-alpha from the optical spectrum is shown as a dashed vertical line. The flux in this detection is $2.5 \times 10^{-16}$ erg s$^{-1}$ cm$^{-2}$ (8 sigma in 3 hours on a 4.2-m telescope).

## 6. MULTI-IFU SPECTROSCOPY

We are currently building a multi-IFU fibre feed for CIRPASS. Figure 10 shows the format of one of the deployable IFUs. Initially we will have 16 of these and eventually we will have 32 (once we have rearranged the spectrograph optics to utilize the full slit length). The fibres have 100 micron cores. The design of each IFU is similar to the large IFU described above. However, to make them more deployable the lens arrays are made from physically smaller lenses. The system will use a plug-plate to position the IFUs in the field. On an 8m telescope such as the VLT the spatial sampling will be ~0.4 arcsec/lens and the FOV of each IFU will be ~3.0×2.0 arcsec. Each IFU can be independently rotated to align it with its target galaxy.

The example shown in figure 10 is for a sample of field galaxies at z~1. Another way in which intend to use our deployable IFUs is to make spatially resolved studies of the ~10 lensed images of the same background galaxy as produced by strong-lensing in galaxy clusters. CIRPASS will be the optimal way of exploiting these powerful gravitational telescopes, allowing very detailed studies of individual galaxies at very early epochs.

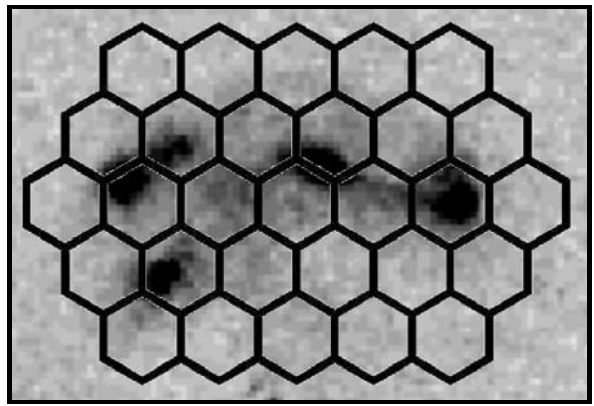

Fig 10. The overlay footprint of one of our 29-element CIRPASS deployable IFUs on top of a z~1 star-forming galaxy. This image comes from HST/ACS B-band and is typical of the spatially extended sample we intend to observe. Note the HII regions – the prominent star-forming knots of emission, which will produce detectable H-alpha.

## 7. CONCLUSIONS

CIRPASS compares very well against other existing and planned NIR spectrographs. At the time of writing it is the only operational spectrograph using a Hawaii-II array. There are very few NIR multi-object spectrographs available or planned and there will not be a NIR multi-IFU available anywhere until 2009 at the earliest.

Its apparent weakness is its throughput, which on the face of it, does not compare well with simpler spectrograph designs such as that of ISAAC for example. However, despite this, we achieve very competitive sensitivities because we have high spectral resolution to get cleanly between the OH lines, excellent instrument stability for long integrations, no slit losses in the IFU mode, a large multiplex gain in the MOS and multi-IFU modes and very low instrument/detector background. We therefore believe that for a significant number of important observing programs CIRPASS is currently the best instrument in the world.

## ACKNOWLEDGMENTS

Firstly, we would like to thank the Raymond and Beverly Sackler Foundation for funding a substantial fraction of this work. Thanks are also very much due to PPARC for funding the construction of CIRPASS and their continued support of its operation. Ben Metcalf, Lexi Moustakas, Stuart Ryder, Gavin Dalton, Ian Lewis, Mark McCaughrean and Simon Hodgkin have all contributed significantly to planning and carrying out our science programs. We thank the AAO for the kind loan of their FOCAP system and two diffraction gratings. Finally, the staff of the Gemini Observatory, the AAO, and the Isaac Newton Group in La Palma have been tremendously helpful and supportive and a pleasure to work with.